Research article

# Evidence of phonon-assisted tunnelling in electrical conduction through DNA molecules

Povilas Pipinys and Antanas Kiveris*

Address: Department of Physics, Vilnius Pedagogical University, Studentu 39, LT 08106 Vilnius, Lithuania

Email: Povilas Pipinys - studsk@vpu.lt; Antanas Kiveris* - studsk@vpu.lt

* Corresponding author

## Abstract

We propose a phonon-assisted tunnelling model for explanation of conductivity dependence on temperature and temperature-dependent *I-V* characteristics in deoxyribonucleic acid (DNA) molecules. The capability of this model for explanation of conductivity peculiarities in DNA is illustrated by comparison of the temperature dependent *I-V* data extracted from some articles with tunnelling rate dependences on temperature and field strength computed according to the phonon-assisted tunnelling theory.

## Introduction

Conductance properties of DNA have recently attracted a lively interest for theoreticians as well as experimenters [1]. Understanding the charge carriers transfer mechanism along DNA double helix is important for possible applications of DNA molecules in nanoelectronic circuit technology [1-6].

Direct conductivity measurements have shown a very wide range of conducting properties ranging from no conduction [5,7,8] to a good linear conductor [2,9,10], while in other experiments semiconducting conductivity behaviour emerges [3,6,11-14].

The wide range of charge transport behaviour seems to arise from different experimental conditions in which the measurements are carried out. These include the nature of the devices used to measure the conductivity, the sequence and length of the DNA, the type of contacts, the environment in different experiments, etc., all can greatly effects the conductivity of the DNA molecules. For instance, Kasumov et al. [4] have shown that strongly deformed DNA molecules deposited on a substrate, whose thickness is less than half the native thickness of the molecule, are insulating, whereas molecules keeping their native thickness are conducting down to very low temperature with a non-ohmic behaviour characteristics of a one dimensional (1D) conductor.

Extensive experimental and theoretical work over the past decade has led to substantial clarification of charge transport mechanisms in DNA. The dominant mechanisms appear to be short-range quantum mechanical tunnelling [14-18] or long-range thermally activated hopping [10,13,19-24]. But these mechanisms are not capable to explain all the field- and temperature-behaviour of experimental data associated with conduction of the DNA molecules. Indeed, the hopping models confronted with difficulties in explaining the observed strong conductivity dependence on the temperature along $\lambda$-DNA double helix at high temperatures and a very week dependence at low temperatures [9]. The tunnelling mechanism was excluded in the case of temperature-dependent results [9,11].

We affirm that in many cases the temperature-dependent conductivity of DNA molecules could be explained by the quantum mechanical tunnelling theories in which the impact of phonons on tunnelling rate is included. In the event, "pure" tunnelling can be observed at low temperatures when the vibrations modes of the molecule are frozen. At moderate temperatures the input of phonon energy to the process of tunnelling must be taken into account and contemporary phonon-assisted tunnelling theories (PhAT) [25-27] realise this.

Recently, it has been shown that the PhAT describes well not only the nonlinear *I-V* curves, but also the temperature-dependent conductivity in conducting polymers [28,29]. Therefore, we invoke the PhAT theory to describe some the temperature-dependent experimental data on electrical transport through DNA molecules presented by other authors.

## Model and comparison with experimental data

We suggest that the thermoactivated current through the DNA molecules is caused by the charge carriers released from localised states located between HOMO and LUMO levels of DNA ones [11]. In the *dc* case, the said levels are continuously filled from the electrode. Assuming that the electrons are released from these states due to phonon-assisted tunnelling, we will compare the current (the same as the conductance) dependence on the temperature and field strength with the tunnelling rate dependence on these parameters, computed using the PhAT theory. For this purpose we explore the equation (18) in [27] derived by Makram-Ebeid and Lannoo for the phonon-assisted tunnelling of the electrons from the impurity centre. Taking into consideration the fact that this theory has been evaluated using the Condon approximation, it is more suitable for the molecular structures than other ones. For the tunnelling rate dependence on field strength $E$ and temperature $T$ this theory gives:

$$W(E,T) = \sum_{p=-p_o}^{+p_o} RW_e(\varepsilon_T + p\hbar\omega), \qquad (1)$$

where

$$R = \exp\left(\frac{p\hbar w}{2k_BT} - S\,\mathrm{cth}\,\frac{\hbar w}{2k_BT}\right)I_p\left(\frac{S}{\mathrm{sh}(\hbar w/2k_BT)}\right), \qquad (2)$$

$$W_e(\varepsilon_T) = \left(\frac{2eF}{(2m^*\varepsilon_T)^{1/2}}\right)\exp\left[-4\frac{(2m^*)^{1/2}}{3e\hbar E}\right]\varepsilon_T^{3/2}. \qquad (3)$$

Here $p_o = \varepsilon_T/\hbar\omega$, $\hbar\omega$ is the phonon energy, $\varepsilon_T$ is the centre depth, $I_P$ is the modified Bessel function and $S$ is the Huang-Rhys coupling constant.

In the first instance, the comparison of the *I-V* data measured in the temperature range from 43 K to 294 K for poly(dA)-poly(dT) DNA molecules (from Fig 2a in [11]) with theoretical *W(E,T)* dependences is presented in Fig. 1. The *W(E,T)* dependences were calculated using centre depth $\varepsilon_T$ to be 0.21 eV (the value is slightly higher than the value of the activation energy estimated in Table 1, which is offered in Ref. [11]), and for the electron effective mass the value of 1.5 $m_e$ [30] was used. For the phonon energy, the value of 43 meV was selected. This value is similar to the value of 348 cm$^{-1}$ used for the calculation of the DNA IR active modes in [31]. The coupling constant *S* was chosen in order to get the best fit of the experimental data with the calculated dependences on the assumption that the field strength for tunnelling is proportional to the applied voltage. As is seen in Fig. 1, the theoretical *W(E,T)* dependences fit well with the experimental data for entire range of the measured temperatures. The field strength for theoretical curves ranges from 0.16 MV/m to 500 MV/m, which is close to the field strength estimated from the sample thickness (about 20 nm).

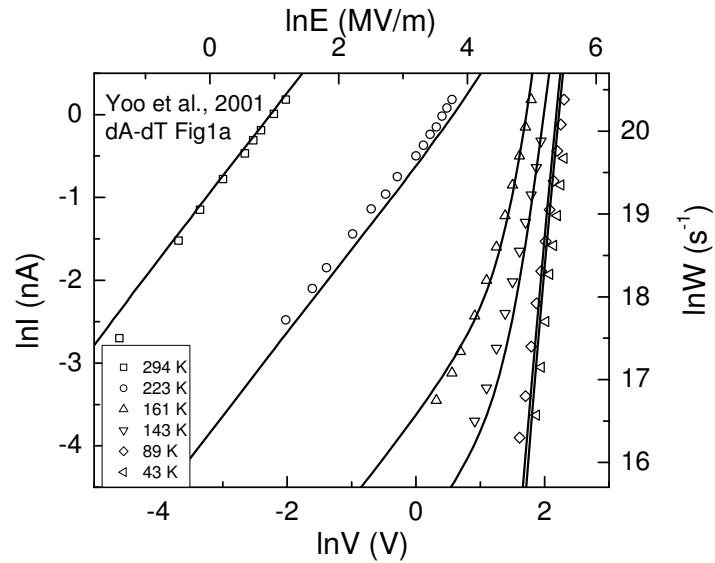

**Figure 1**
Comparison of *I – V* dependences in poly(dA)-poly(dT) DNA molecules extracted from figure 2(a) in [11] (symbols) with theoretical *W (E,T)* against *E* dependences (solid curves) calculated for the same *T* as in the experiment (from top to bottom) using the following parameters: $\varepsilon_T$ = 0.21 eV, $m^*$ = 1.5 $m_e$, $\hbar\omega$ = 43 meV and $S$ = 12.

The judgment on the carriers transfer mechanism is often carried out considering the conductance dependence on the temperature. The conductance measured by Yoo et al. for poly(dA)-poly(dT) was strongly temperature-dependent around room temperature and slightly temperature-dependent at low temperatures [11]. The authors explain such dependence within small polaron hopping model. We note that the *W (E,T)* versus *E* characteristics at temperatures below 100 K are weakly dependent on the temperature, and this feature is in excellent agreement with the experimental observation. This circumstance is also evident in Fig. 2 (solid lines) from the plot of ln *W (E,T)* versus *1/T* calculated using the same parameters as in Fig. 1 and for *E* = 135 MV/m. The symbols in Fig. 2 represent the experimental data extracted from figure 2(c) in [11]. One can see that the theoretical dependences of the phonon-assisted tunnelling rate are in good agreement with the experimental data.

Tran et al. [9] using the resonant cavity technique studied conductivity and its temperature dependence along the *λ*-DNA double helix at microwave frequencies. They observed similar behaviour of the conductivity on the temperature as in [11], i.e. a strong temperature dependence of conductivity at high temperatures, whereas at low temperatures the conductivity in *λ*-DNA exhibits a very weak temperature dependence. They explain the temperature-dependent conductivity suggesting two transport mechanisms, i.e. ionic conduction at low temperatures and temperature driven hopping transport processes at high temperatures. The underlying physics of the weak temperature dependence at low temperatures was not understood. The calculation in [32] has also shown that transport through DNA does not have a purely hopping character.

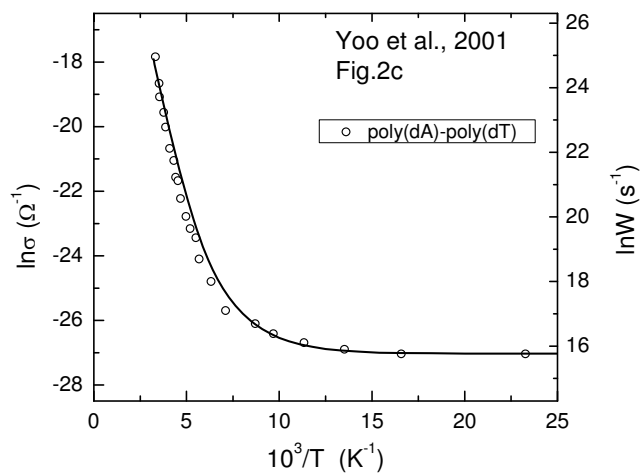

**Figure 2**
Experimental *σ (T)* against *1/T* dependence for poly(dA)-poly(dT) DNA molecules extracted from figure 2 (c) in [11] (symbols) fitted to *W (E,T)* against *1/T* dependence, calculated using the same parameters as in Fig. 1.

In Fig. 3 the experimental data extracted from Fig. 3 in [9] for *λ*-DNA (symbols) fitted to the PhAT rate dependences on *1/T* for the *λ*-DNA in buffer (solid curve) and for the dray *λ*-DNA (dashed curve), are depicted. In this case the theoretical *W(T)* dependences reflect also the experimental data well.

## Conclusion

In conclusion, the PhAT model is able to explain the peculiarities of field- and temperature-dependent current observed in DNA molecules in a wide region of the electric field strength. A strong temperature dependence of conductivity observed at high temperatures and a very weak temperature dependence at low temperatures of DNA molecules is comprehensible in the framework of the PhAT model. It is worth to note that the *W(E,T)* dependences at both low and high temperatures are calculated using the same set of parameters, i.e. $\varepsilon_T$, $m^*$, *T*, *E*, $\hbar\omega$ and *S*. From these ones only the Huang-Rhys coupling constant S is the fitting parameter estimated from the best fitting of the experimental data and theory. Other parameters are known from experiments or from literary sources. An advantage of the PhAT model over the other models used is the possibility to describe the behaviour of the *I-V* data measured at different temperatures with the same set of parameters characterizing the material.

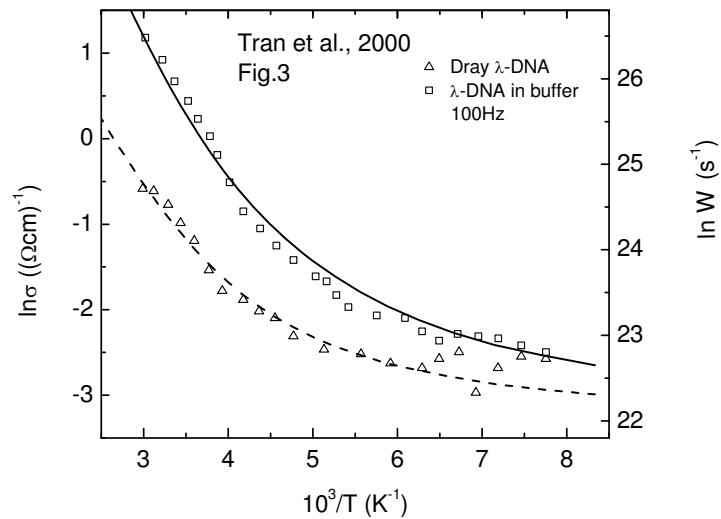

**Figure 3**
Experimental $\sigma(T)$ against *1/T* dependences for the *λ*-DNA in buffer and for the dray *λ*-DNA extracted from figure 3 in [9] (symbols) fitted to theoretical *W(E,T)* against *1/T* dependences calculated for the following parameters: $\varepsilon_T$ = 0.21 eV, *E* = 290 MV/m, *S* = 12 (solid line) and for parameters $\varepsilon_T$ = 0.17 eV, *E* = 315 MV/m, *S* = 8 (dashed line).

Thus, the PhAT mechanism, in some cases, could be dominant in the conductance of the DNA molecules.